\documentclass[nofootinbib,prx,twocolumn,showpacs,superscriptaddress,notitlepage,superscriptaddress,amsmath]{revtex4-2}
\usepackage[T1]{fontenc}

\usepackage{lipsum}  
\usepackage{dsfont}
\usepackage{amsmath}
\usepackage{braket}
\usepackage{amssymb}
\usepackage{amsthm}
\usepackage{algpseudocode}
\usepackage{algorithm}
\usepackage{siunitx}
\usepackage{amsfonts}
\usepackage{comment}
\usepackage[normalem]{ulem}
\usepackage{graphicx}
\usepackage{color,framed}
\usepackage{hyperref}
\usepackage{enumerate}
\usepackage{slashed}
\usepackage{url}
\usepackage{bbm}
\usepackage{chngcntr}
\counterwithout{equation}{section}
\usepackage{tikz,pgfplots}
\usepackage{pgfplotstable}
\usepackage{siunitx}
\usepackage{comment}
\usepackage{graphicx}

\makeatletter
\newcommand{\setlabel}[1]{\edef\@currentlabel{#1}\label}
\makeatother

\usepgfplotslibrary{fillbetween}
\hypersetup{
    colorlinks=true,
    linkcolor=blue,
    filecolor=magenta,      
    urlcolor=brown,
}

\hypersetup{
    colorlinks=true, 
    linktoc=all,     
    linkcolor=blue,  
}
\def \beq {\begin{equation}}
\def \eeq {\end{equation}}
\def \beqa {\begin{eqnarray}}
\def \eeqa {\end{eqnarray}}
\def \bseq {\begin{subequations}}
\def \eseq {\end{subequations}}

\newcommand{\D}[1]{\text{d}#1}

\pgfplotsset{compat=1.18}
\begin{document}

\title{Less precise but less noisy:\\ local circuits for momentum-space state preparation and measurement}

\author{Etienne Granet}
\email{etienne.granet@quantinuum.com}
\author{Henrik Dreyer}
\affiliation{Quantinuum, Leopoldstrasse 180, 80804 Munich, Germany}
\date{\today}

\begin{abstract}

Quantum algorithms are usually optimized for gate count or circuit depth. We find on Quantinuum System Model H2 quantum computer that for a tight-binding chain ground state preparation, there is a system size $N$ beyond which the adiabatic evolution reaches significantly lower energies than the Fermionic Fourier Transform (FFT), with the same number of gates, and with the same circuit depth. We attribute this high noise sensitivity of the FFT to its high precision, being able to distinguish momenta by $1/N$. This high resolution in momentum space requires long-range couplings in real space, which propagates errors faster. In contrast, although local and physical circuits such as the adiabatic evolution have a coarser momentum resolution, they also propagate errors more slowly. For physical applications, high momentum resolution is rarely required and is often worth trading for low noise sensitivity. We also introduce a momentum measurement scheme that although less precise than FFT, is less costly and less noisy. We show that it achieves better performance than FFT for spectral function measurement on Quantinuum System Model H2 quantum computer. Our work emphasizes the importance of reducing the noise sensitivity of quantum algorithms, beyond the number of gates or circuit depth.


\end{abstract}

\maketitle

\section{Introduction} 
One of the earliest applications of quantum computers is the simulation of many-body quantum physical systems \cite{miessen2023quantum,alexeev2024quantum}, with recent progress across multiple quantum computing architectures \cite{kim2023evidence,haghshenas2025digital,abanin2025observation,alam2025programmable,evered2025probing,granet2025superconducting}. This finds applications in material science, chemistry, condensed-matter physics and many other domains. To reach lower temperatures or longer evolution times, hardware noise or the effects of hardware noise must be reduced in some way. To that end, circuits are almost always optimized for either gate count or circuit depth.

\begin{figure}
    \centering
    \includegraphics[width=1\linewidth]{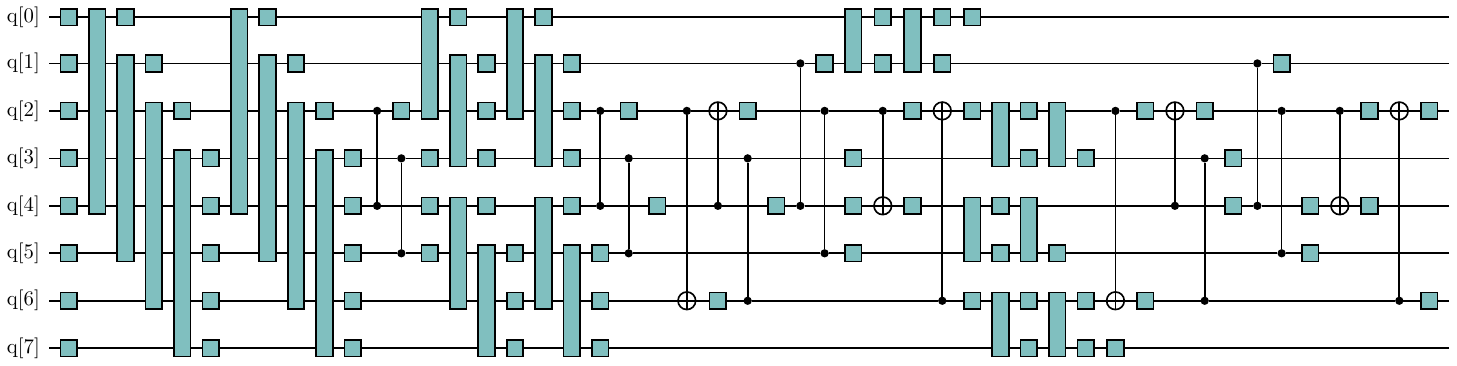}

    \vspace{0.15cm}
    \includegraphics[width=1\linewidth]{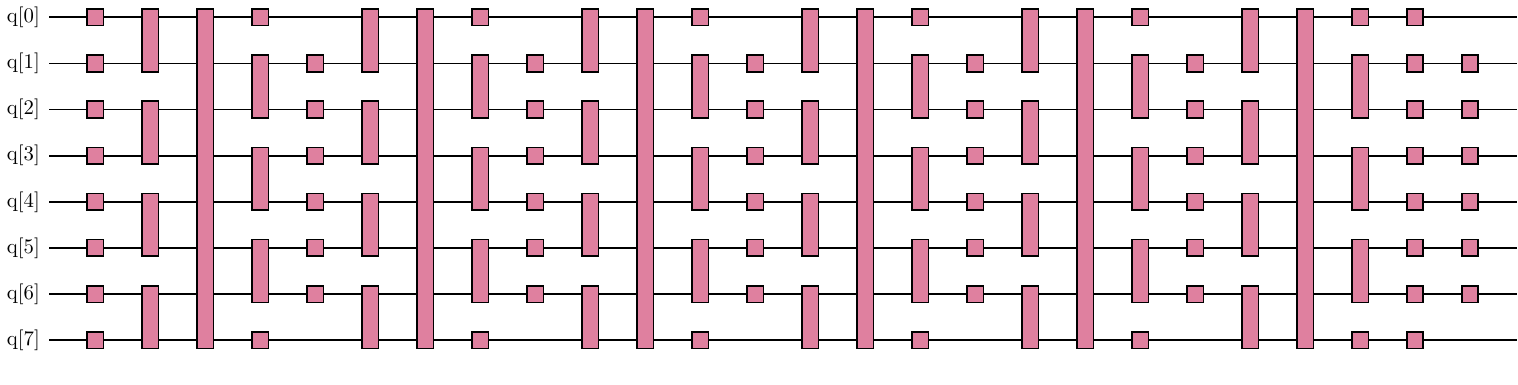}
    
    \vspace{0.15cm}
    \includegraphics[width=0.7\linewidth]{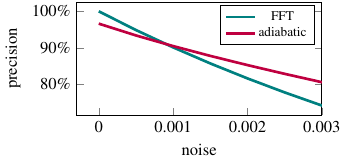}
    
    \caption{Sketch of two circuits with approximately same number of gates that prepare the ground state of a tight-binding chain in size $N=8$ with good precision. While the fermionic Fourier transform (top, green) is more precise in the absence of noise, the adiabatic evolution (bottom, red) sees its precision degrade less quickly with noise. The plot displays the numerically computed relative energy compared to the ground state energy in size $N=48$, for approximately the same number of two-qubit gates, as a function of depolarizing noise rate after every two-qubit gate. }
    \label{fig:conceptual}
\end{figure}

The main objective of this paper is to argue that \emph{noise sensitivity} of the circuit should also be taken into account in this optimization. Different circuits can have very different sensitivity to hardware noise \cite{xue2021effects,fontana2021evaluating,schiffer2024quantum,granet2025dilution,chertkov2026robustness,gonzalez2026stability,vazquez2023well}. Even though two circuits achieve the same goal with the same number of gates or the same circuit depth, their implementation on actual hardware can give very dissimilar results. We will demonstrate this idea in the context of fermionic simulation, taking as two paradigmatic routines the Fermionic Fourier Transform (FFT) \cite{verstraete2009quantum,ferris2014fourier,babbush2018low} and Hamiltonian simulation \cite{lloyd1996universal}, whose circuits are sketched in Fig \ref{fig:conceptual} for a small system size.

The FFT is an operator that allows one to switch between real space and momentum space. It is a "fully digital" circuit that gives exquisite control of momentum modes and whose compilation has been optimized in several works \cite{kivlichan2018quantum,maskara2025fast,constantinides2025low,aigner2026fermion}. It appears in multiple contexts, including state preparation \cite{verstraete2009quantum,velury2025quasiparticle}, quantum dynamics \cite{kivlichan2018quantum,babbush2018low,babbush2018encoding,childs2021theory,campbell2022early,schuckert2024fault,li2026asymptotically,aigner2026fermion}, and momentum mode measurement \cite{babbush2018low,tong2021fast,velury2025quasiparticle,granet2026spectral}. On the other hand, Hamiltonian simulation is more "analog" in spirit as it implements a physical process, and is only approximated by gate-based quantum computers. It can be used instead of FFT for some applications, but with less precision, deeper circuit and more gates. We will show that in spite of this, Hamiltonian simulation used for ground state preparation achieves significantly lower energies than FFT on Quantinuum H2-2 quantum computer, beyond a system size $N\approx 20$, even with \emph{same number of gates} in the circuit or \emph{same depth}. We attribute this better performance on noisy hardware to the lower precision in momentum space of Hamiltonian simulation techniques compared to FFT. Finer precision in momentum space means longer-range gates in real space, which propagate errors faster. We emphasize again that this better performance of the adiabatic evolution at preparing low-energy states is due to the presence of gate noise: on a perfect hardware, the FFT will always be cheaper and prepare the perfect ground state. This observation applies to today's NISQ computers \cite{preskill2018quantum}, but also beyond the NISQ era, whenever a machine is pushed to the limits of its capabilities and logical errors cannot be neglected.

Beyond this observation and analytical explanation, we propose an alternative to FFT for momentum measurement. We implement our momentum measurement scheme on Quantinuum H2-2 hardware for a spectral function measurement problem, and observe significant improvement over the FFT.

\section{Noise sensitivity of FFT and adiabatic evolution \label{sec:noisesensitive}}

\subsection{Setup and objective}
We first directly present evidence for the main message of this paper, namely there exist alternatives to FFT that perform better on actual quantum hardware for a given task, even though they are less efficient or less precise on ideal noiseless hardware. The example that we consider is that of ground state preparation of a tight-binding chain
\begin{equation}\label{tb}
    H_{\rm tb}=-\sum_{j=1}^N c_j^\dagger c_{j+1}+c^\dagger_{j+1}c_j\,,
\end{equation}
with antiperiodic boundary conditions $c_{N+1}=-c_1$, and where we will always assume $N$ even. This consists in preparing a state of the form
\begin{equation}\label{GS}
    c^\dagger(k_1)...c^\dagger(k_n)|0\rangle\,,
\end{equation}
where the momenta $k_i$ are all the momenta available between some maximal momenta $-k_{*}$ and $k_{*}$. Here, $c(k)=\frac{1}{\sqrt{N}}\sum_{j=1}^N c_j e^{ijk}$ denotes the Fourier transform of the fermionic operator $c_j$ in real space, with $k=\frac{2\pi(n+1/2)}{N}$ for $n=0,...,N-1$. The objective is to prepare this state on hardware with two completely different circuits with approximately same number of gates or same circuit depth, but with very different accuracy on the value of the ground state energy.

\subsection{FFT and adiabatic evolution \label{sec:adiabatic}}
The state \eqref{GS} can be exactly prepared on an ideal noiseless hardware by applying a FFT on a product state $|0...01...10...0\rangle$ where the $1$s correspond to the momenta between $-k_{*}$ and $k_{*}$. Indeed, the FFT is an operator that maps real-space fermions to momentum-space fermions, namely
\begin{equation}
     {\rm FFT}^\dagger c_j {\rm FFT} = \frac{1}{\sqrt{N}}\sum_{\ell=1}^{N}c_\ell e^{\frac{2i\pi}{N}\ell(j+1/2)}\,.
\end{equation}
Here, we have shifted the momenta by $1/2$ because of antiperiodic boundary conditions, which differs from the usual FFT just by single-qubit $Z$ rotations on top. With this, we have
\begin{equation}
\begin{aligned}
    c^\dagger(-k_*)...c^\dagger(k_*)|0...0\rangle&= {\rm FFT} c^\dagger_{-n_*}...c^\dagger_{n_*-1} {\rm FFT}^\dagger |0...0\rangle\\
    &= {\rm FFT} |0...011...110...0\rangle\,,
\end{aligned}
\end{equation}
where the integers $-n_*,-n_*+1,...,n_*-1$ are such that $k_*=\frac{2\pi (n_*-1/2)}{N}$, with indices understood modulo $N$.

Another completely different approach to ground state preparation is the \emph{adiabatic evolution}. When preparing the state of the system in the ground state of a Hamiltonian $H_0$, and time-evolving the system with a slowly-varying time-dependent Hamiltonian that interpolates between $H_0$ and $H_{\rm tb}$ in \eqref{tb}, one prepares an approximation of the ground state of $H_{\rm tb}$, provided there is no gap closing along the path. We consider the following initial Hamiltonian
\begin{equation}\label{h0cc}
    H_{0}=-\sum_{j=1}^N c_j^\dagger c_{j+1}+c^\dagger_{j+1}c_j+c^\dagger_{j+1}c_j^\dagger +c_jc_{j+1}\,,
\end{equation}
with antiperiodic boundary conditions. With a Jordan-Wigner encoding in the sector with an even number of particles $\prod_{j=1}^N Z_j=1$, the Hamiltonians $H_0$ and $H_{\rm tb}$ become
\begin{equation}
    \begin{aligned}
        H_{0}&=-\sum_{j=1}^N X_j X_{j+1}\\
        H_{\rm tb}&=-\frac{1}{2}\sum_{j=1}^N X_j X_{j+1}+Y_j Y_{j+1}\,.
    \end{aligned}
\end{equation}
We will study analytically the efficiency of this adiabatic evolution in Section \ref{seceff} below. We are going to show that beyond a certain system size, the adiabatic evolution achieves better energies than the FFT on actual hardware, with the same number of gates.

\subsection{Hardware implementation}
\begin{figure*}
    \centering
    \includegraphics[width=0.8\linewidth]{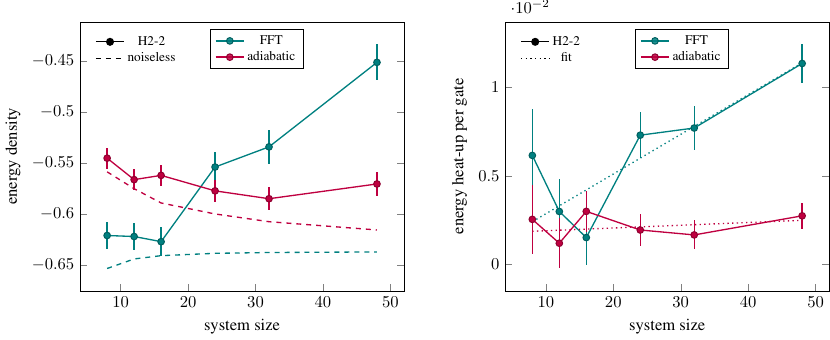}
    \caption{\emph{Left panel:} Energy density as a function of system size, measured on Quantinuum H2-2 quantum computer (solid lines) and noiseless values (dashed lines), using a FFT (teal) and an adiabatic evolution (purple). We choose the number of steps in the adiabatic evolution so as to have approximately the same number of two-qubit gates as the FFT for every system size. Error bars indicate one standard deviation. \emph{Right panel:} Energy increase per two-qubit gate due to noise, as a function of system size measured on Quantinuum H2-2 quantum computer, from the same data as left panel. The dotted lines show a linear fit of the data.}
    \label{fig:energy_comparison}
\end{figure*}

We present a hardware implementation on Quantinuum H2-2 quantum computer \cite{moses2023race}. For different values of $N$, we implement the FFT with unitary circuits with a depth scaling as $\log^2 N$ (see Ref \cite{constantinides2025low,granet2026spectral}). The adiabatic evolution for $N_{\rm Trott}$ Trotter steps is implemented as
\begin{equation}
    U=\prod_{t=1}^{N_{\rm Trott}} e^{i x_t \sum_{j=1}^N X_j X_{j+1}}e^{i y_t \sum_{j=1}^N Y_j Y_{j+1}}\,,
\end{equation}
with $y_t/x_t$ going from $0$ to $1$ along the path, applied on $|+...+\rangle$. Specifically, we choose $x_t=d_t$ and $y_t=d_t \frac{t-0.5}{N_{\rm Trott}}$, and $d_t=0.4\sqrt{1-\frac{t-0.5}{N_{\rm Trott}}}$. This is a linear schedule with a time-dependent Trotter step that is empirically observed to perform well. In principle, the adiabatic evolution must be carried out from
\begin{equation}\label{gs}
\frac{|+...+\rangle+|-...-\rangle}{\sqrt{2}}
\end{equation}
in order to give the ground state of $H$ in the even-particle sector. However, this GHZ state preparation can always be replaced by starting from $|+...+\rangle$ and post-selecting shots with an even number of particles, since the operator $U$ preserves the parity of the number of particles. The discard rate would be $50\%$. Even without this post-selection, the energy difference obtained is tiny since the difference of ground state energy for $H_{\rm tb}$ with even or odd number of particles scales as $1/N$. For the practicality of our approach, we will thus fix the initial state to be $|+...+\rangle$. We set the number of Trotter steps $N_{\rm Trott}$ in the adiabatic evolution so that for every $N$ there is (approximately) the same number of two-qubit gates in the FFT and in the adiabatic evolution circuit. Specifically, for every $N$ we take the smallest integer $N_{\rm Trott}$ such that the adiabatic evolution has more two-qubit gates than the FFT, and the precise two-qubit gate counts are given in Appendix \ref{appgatecount}. We then measure the final energy for both circuits, and report the raw values in Fig \ref{fig:energy_comparison}. 

We observe that at small system sizes, the FFT reaches lower energies than the adiabatic evolution. However, at larger system sizes $N\gtrapprox 20$, there is a crossing between the two curves and the adiabatic evolution reaches lower energies than the FFT. The energy error strongly increases with system size for the FFT, whereas at large system sizes it increases only mildly (or even is constant) for the adiabatic evolution. We emphasize again that for a same system size, the two circuits, FFT and adiabatic, have approximately the same number of two-qubit gates. In terms of circuit depth, the adiabatic circuits are shallower than the FFT since they are denser. At $N=48$, the two-qubit gate circuit depth of the adiabatic circuit is around twice smaller than the FFT circuit. We also ran on hardware at $N=48$ an adiabatic evolution with approximately same circuit depth as the FFT, which corresponds to around twice more two-qubit gates. We obtain an energy $-0.577 \pm 0.017$, which is even lower than the result for same number of two-qubit gates plotted in Fig \ref{fig:energy_comparison}. This shows that in size $N=48$, the adiabatic evolution always performs better than the FFT, whether we impose the same number of two-qubit gates or the same circuit depth, even though the FFT performs better in the noiseless case.

In the right panel of Fig \ref{fig:energy_comparison}, we then plot the difference between measured energy and a noiseless simulation of the circuit, divided by the number of two-qubit gates in the circuit. This gives an effective energy heat-up per two-qubit gate due to hardware noise. We observe that for the adiabatic evolution, the amount of heating per gate is roughly constant with system size, while for the FFT, it grows linearly with system size. The linear growth with system size means that every error affects a number of modes that is $\mathcal{O}(N)$, whereas the constant behaviour of the adiabatic evolution indicates that every error impacts the energy by just $\mathcal{O}(1)$. This exactly corresponds to the presence of dilution of error for the adiabatic evolution \cite{granet2025dilution}, typical of Hamiltonian simulation circuits, and to the absence of it for the FFT, typical of global, "digital" circuits. For Hamiltonian simulation, errors typically get "diluted" and impact local observables by only $\mathcal{O}(1/N)$, against $\mathcal{O}(1)$ for general circuits.

\subsection{Efficiency of adiabatic evolution \label{seceff}}
\subsubsection{Generalities}
The objective of this section is to study analytically the efficiency of the adiabatic evolution for the tight-binding chain ground state preparation. We are going to show that the energy obtained as a function of the adiabatic time displays two regimes, an intermediate-time regime with a polynomial decrease of energy, and a late-time regime with an exponential decrease. Then we will study the effect of gate noise on the adiabatic evolution and explain, as observed on hardware, that there is a noise-dependent system size $N^*$ beyond which the adiabatic evolution performs better than the FFT. We will also study the momentum content of the state obtained with a finite adiabatic time.

\subsubsection{Scaling of the energy with the adiabatic time}
The efficiency of the adiabatic evolution depends on two aspects, first on the smoothness of the path, and second on the smallest gap along the adiabatic path \cite{jansen2007bounds,elgart2012note,cheung2011improved,mackenzie2006perturbative}. In our case, the efficiency of the adiabatic evolution can be made more precise with the following reasoning.

As is well-known, the Hamiltonians $\sum_j X_j X_{j+1}$ and $\sum_{j}Y_j Y_{j+1}$ can be written in terms of quadratic functions of fermionic operators $c_j$, as in \eqref{h0cc}. When written in momentum space, they read
\begin{equation}
\begin{aligned}
&\sum_{j=1}^N X_j X_{j+1}\,,\sum_{j=1}^NY_j Y_{j+1}=\\
    &\sum_{k} 2\cos k c^\dagger(k) c(k)\pm i\sin k (c(k) c(-k)-c^\dagger(-k)c^\dagger(k))\,,
\end{aligned}
\end{equation}
with the $\pm$ sign corresponding to the $X/Y$ Hamiltonians. These Hamiltonians preserve the following structure of states
\begin{equation}\label{coherent}
    |\psi\rangle=\prod_{k>0} (A_k +B_k c^\dagger(-k) c^\dagger(k))|0\rangle\,,
\end{equation}
for complex numbers $A_k,B_k$, where the product is over $k=\frac{2\pi(n+0.5)}{N}$ for $n=0,...,N/2-1$. Namely, the time evolution of this state with $\sum_j X_j X_{j+1}$ or $\sum_j Y_j Y_{j+1}$ can be written in the same form \eqref{coherent} with some $A_k',B_k'$ that depend only on $A_k,B_k$ with the same $k$. Since the initial ground state of $H_0$ is of the form \eqref{coherent} \cite{calabrese2012quantum}, it follows that the adiabatic evolution presented in Section \ref{sec:adiabatic} factorizes into a tensor product of $N/2$ adiabatic evolutions within $2$-dimensional Hilbert spaces, spanned by the two states where the momenta $k$ and $-k$ are either simultaneously occupied or empty. In this subspace, we can identify $c^\dagger(k) c(k)+c^\dagger(-k) c(-k)$ with $1-Z$, $c^\dagger(-k)c^\dagger(k)+c(k)c(-k)$ with $X$ and $-i(c^\dagger(-k)c^\dagger(k)-c(k)c(-k))$ with $Y$. Neglecting Trotter error, the effective Hamiltonians $H_k$ within these subspaces are the one-qubit Hamiltonians
\begin{equation}\label{hkt}
    H_k(t)=[2\cos k] I-[2\cos k] Z +[2(1-t/T)\sin k] Y\,.
\end{equation}
Here, $T$ denotes the total time evolution, and $t$ parametrizes the interpolation between $H_0$ and $H_{\rm tb}$. We note that Trotterization that alternates the $XX$ and $YY$ Hamiltonians does not break this factorized structure and the same reasoning would apply, although analytical expressions would be less simple.

For a smooth path with all derivatives vanishing at the beginning and at the end, each of the adiabatic evolutions with $H_k(t)$ converges exponentially fast to the ground state of the $2\times 2$ systems on which it is defined. Each of these convergences occurs with their own time scale $T_k$, which depends on the smallest gap in each of these sectors separately. Since the energy gap at the end of the path is $4\cos k$, it follows that the energy density above the ground state scales as
\begin{equation}\label{deltae}
    \delta e(T)=\frac{1}{N}\sum_{k>0}4 |\cos (k)| e^{-T/T_k}\,.
\end{equation}
In the late asymptotic regime in $T$, only $k^*$ with the largest $T_k$ remains, and we have the leading $T\to\infty$ behaviour
\begin{equation}\label{latetime}
    \delta e(T)\underset{T\to\infty}{\sim} \frac{4|\cos (k^*)|}{N}e^{-T/T_{k^*}}\,.
\end{equation}
This asymptotic result corresponds to the standard result about adiabatic evolution being controlled by the smallest gap along the path. 

To go further, let us determine some precise values for $T_k$. Our reasoning starts from the Landau-Zener model \cite{landau1932theorie,zener1932non}, that is a time-dependent Hamiltonian for one qubit $H_{\rm LZ}(t)=vt Z+\Delta X$, where $v,\Delta$ are parameters. It is known that the probability that the state $|0\rangle$ remains in the state $|0\rangle$ after evolving with $H(t)$ from $t=-\infty$ to $t=\infty$ is $e^{-\pi \Delta^2/v}$. We would like to equate this value with $e^{-T/T_k}$ for some $T_k$. Since the asymptotic behaviour is controlled by the behaviour of the Hamiltonian near the minimal gap, we determine the parameters $v$ and $\Delta$ by identifying $H_{\rm LZ}$ with $H_k$ near the minimal gap. We identify the coefficient of $X$ in $H_{\rm LZ}$ with that of $Z$ in $H_k$, and the derivative with time of the coefficient of $Z$ in $H_{\rm LZ}$ with that of $Y$ in $H_k$. We deduce $\Delta=2\cos k$ and $v=\frac{2\sin k}{T}$. From this we find
\begin{equation}\label{tk}
    T_k=\frac{\sin k}{2\pi \cos^2 k}\,.
\end{equation}
The value of $k$ that maximizes $T_k$ is the momentum closest to $\pi/2$ in finite size $N$. Then, since $\cos k^*$ is of order $1/N$, we have $T_{k^*}=\mathcal{O}(N^2)$, which is the standard scaling of inverse minimal gap square for the adiabatic evolution. The energy then scales as
\begin{equation}\label{expscaling}
    \delta e(T)\underset{T\to\infty}{\sim} \frac{a}{N^2}e^{-b T/N^2}\,,
\end{equation}
with some coefficients $a,b$ of order $1$ in $N$. Once again, this scaling corresponds to the \emph{very late} asymptotic regime $T\to\infty$ at fixed $N$.


Let us now consider instead another regime. We take the limit $N,T\to\infty$ while keeping $T\ll T_{k^*}$. In that case, we can approximate the sum \eqref{deltae} by an integral when $N\to\infty$
\begin{equation}
    \delta e(T)=\frac{2}{\pi}\int_0^\pi |\cos(k)|e^{-T/T_k}\D{k}\,,
\end{equation}
with now an $N$-independent $T_k$ that is a function of the continuous variable $k$. Noting that $T_k$ diverges as $|k-k^*|^{-2}$ near $k^*$, and noting that $k^*=\pi/2$, doing a change of variable $u=T^{1/2}(k-k^*)$, we find
\begin{equation}\label{polyscaling}
    \delta e(T)\sim \frac{1}{\pi^2 T}=\mathcal{O}(T^{-1})\,.
\end{equation}
We see thus that before the late-time exponential asymptotic regime, there is an \emph{intermediate} asymptotic regime for $T\ll T_{k^*}$ where the energy density decreases as a \emph{power law} in $T$. This is an important difference with the exponential scaling \eqref{expscaling} that gives very little energy decrease whenever $T\ll N^2$. This power-law is unrelated to the power-law scaling obtained for an adiabatic evolution when the path is not smooth enough. Here, the power-law comes from the fact that before the late asymptotic regime is reached, the portion of the Hilbert space that is not yet converged decreases, resulting in a decreasing energy even though the evolution is still too fast near the minimal gap. 

Standard results on the adiabatic theorem only apply to the late-time exponential regime that is governed by this smallest gap along the path. However, in practice,  a given required precision  $\delta e$ may be reached within this intermediate polynomial regime, and the exponential regime is never involved. Indeed, since the scaling \eqref{expscaling} occurs for $T\gg N^2$ and the scaling \eqref{polyscaling} for $T\ll N^2$, we have that when targeting a precision $\epsilon$ on $\delta e$, we must take an adiabatic time $T(\epsilon)$ that scales up to prefactors as
\begin{equation}
    T(\epsilon)\sim \begin{cases}
        \frac{1}{\epsilon}\,,\qquad \text{for }\frac{1}{N^2}\ll \epsilon \ll 1\\
        -N^2 \log (N^2 \epsilon)\,,\qquad \text{for } \epsilon \ll \frac{1}{N^2}\,.
    \end{cases}
\end{equation}
For any precision that is less good than $1/N^2$, the standard late asymptotic regime of the adiabatic evolution is actually never reached. In practice, getting a fidelity $\mathcal{O}(1)$ with the ground state is enough to sample shots from the ground state with $\mathcal{O}(1)$ probability, while getting an infidelity that is vanishingly small (which requires entering the late-time exponential asymptotic regime) is not needed.

\subsubsection{Effect of gate noise}
Let us now study the effect of gate noise on these scalings. We assume that every gate comes with a probability of error $p\ll 1$. In the case of Hamiltonian evolution, it is known that errors have an effect on local observables such as the energy density that is only $\mathcal{O}(1/N)$ \cite{granet2025dilution,chertkov2026robustness}. Since there are $\mathcal{O}(NT)$ gates in the circuit, we get an energy density scaling as
\begin{equation}
    \delta e\approx \frac{1}{\pi^2 T}+\frac{2p cT}{{\rm d}t}\,,
\end{equation}
with $c$ a constant of order $1$, and ${\rm d}t$ the Trotter step size. The optimal value of $T$ is attained for 
\begin{equation}
    T^*=\frac{1}{\sqrt{p}}\sqrt{\frac{{\rm d}t}{2c \pi^2}} =\mathcal{O}\left(\frac{1}{\sqrt{p}}\right)\,,
\end{equation}
and the optimal energy density is then
\begin{equation}\label{sqrtp}
    \delta e(T^*)=2\sqrt{p} \sqrt{\frac{2c}{\pi^2 {\rm d}t}}=\mathcal{O}(\sqrt{p})\,.
\end{equation}
Let us now determine the system size $N^*$ at which the adiabatic evolution is more efficient than the FFT, with the same number of gates. For the FFT, dilution of error does not apply, and every error modifies the energy density by $\mathcal{O}(1)$. It follows that the energy density above ground state for the FFT scales as
\begin{equation}
    \delta e_{\rm FFT}= apNT\,,
\end{equation}
for some constant $a>0$, and where we used that the number of gates is the same as in the adiabatic evolution, $NT$, with a $N$-dependence of $T(N)$ that depends on the compilation chosen for the FFT. At $N=N^*$ we thus have
\begin{equation}
    N^*=\frac{1}{a\pi^2 T(N^*)^2 p}+\frac{2c}{a {\rm d}t}\,.
\end{equation}
For the compilation of FFT in $\log^2(N)$ depth, we have $T(N)\sim \log^2N$. It follows that at small noise rate $p\to 0$ we have the scaling
\begin{equation}
    N^*=\mathcal{O}\left(\frac{1}{p \log^2 p}\right)\,.
\end{equation}
For $N\leq N^*$, the FFT yields lower energies, while at larger system sizes $N>N^*$, the adiabatic evolution yields lower energies.

We note that here, the performance of the adiabatic evolution and of the FFT is evaluated through the energy of the prepared state. One could however consider instead the fidelity with the ground state. In that case, since the energy density difference between the first excited state and the ground state scales as $\mathcal{O}(1/N^2)$, one needs with the adiabatic evolution a time $T=\mathcal{O}(N^2)$ to reach a fidelity with the ground state arbitrarily close to $1$. Since for global fidelity there is no dilution of error, the effect of noise is to multiply this fidelity by $e^{-apN^3}$ with some constant $a>0$, assuming $p$ small. For the FFT, this factor due to noise is instead $e^{-bpN \log^2N}$ with some constant $b>0$. At large system sizes, the FFT performs thus always better than the adiabatic evolution on noisy hardware, when taking as a metric the fidelity with the ground state. The better performance of the adiabatic evolution compared to FFT when measuring the energy is directly related to dilution of error occurring for local observables in Hamiltonian simulation \cite{granet2025dilution,chertkov2026robustness}.

\subsubsection{Generalization to other dispersion relations and dimensions}
Let us apply the previous reasoning to more general dispersion relations $\varepsilon(k)$. Namely, we consider an adiabatic evolution between $H_0$ and $\sum_k \varepsilon(k) c^\dagger(k) c(k)$. The adiabatic evolution in the sector with momentum $k,-k$ is again described by the one-qubit Hamiltonian
\begin{equation}
    H_k(t)=\varepsilon(k)\frac{I-Z}{2} +[2(1-t/T)\sin k] Y\,.
\end{equation}
Repeating the arguments, the convergence time for mode $k$ is $T_k=\frac{8\sin k}{\pi\varepsilon(k)^2}$ and in the intermediate regime the energy above the ground state scales as
\begin{equation}
    \delta e(T)\sim \int_0^\pi |\varepsilon(k)| e^{-T/T_k}\D{k}\,.
\end{equation}
Let us assume that the dispersion relation vanishes at $k=k^*$ as $\varepsilon(k)\sim |k-k^*|^z$ with some power $z$. Then in the case where $k^*\neq 0,\pi$ we obtain
\begin{equation}
    \delta e(T)=\mathcal{O}( T^{-\frac{1+z}{2z}})\,.
\end{equation}
For example, for a quadratically vanishing dispersion $z=2$, the energy density decreases more slowly than before, as $T^{-3/4}$. In the case where $k^*=0$ or $k^*=\pi$, we get
\begin{equation}
    \delta e(T)=\mathcal{O}( T^{-\frac{1+z}{2z-1}})\,.
\end{equation}
The effect of noise can be treated simultaneously for these cases by assuming a dependence $\delta e\sim T^{-\gamma}$ for  some $\gamma$. We find, assuming dilution of error, that the optimal adiabatic time scales as
\begin{equation}
    T^*=\mathcal{O}(p^{-1/(1+\gamma)})\,,
\end{equation}
and the energy
\begin{equation}
    \delta e(T^*)=\mathcal{O}(p^{\gamma/(1+\gamma)})\,.
\end{equation}
Then the limiting system size $N^*$ after which the adiabatic evolution reaches lower energies than the FFT with $\log^2$ depth scales as $N^*=\mathcal{O}\left(1/(p \log^{1+\gamma}p)\right)$.

Let us now comment on the case of higher dimension $d>1$. As an example, we will consider the simplest generalization of the 1D case studied above and take
\begin{equation}
    H_{0}=-\sum_{\ell}\sum_{j=1}^L c_{j,\ell}^\dagger c_{j+1,\ell}+c^\dagger_{j+1,\ell}c_{j,\ell}+c^\dagger_{j+1,\ell}c_{j,\ell}^\dagger +c_{j,\ell}c_{j+1,\ell}\,,
\end{equation}
where $(j,\ell)$ denotes the $d$-dimensional coordinate, with $j$ denoting the first coordinate and $\ell$ the $d-1$ remaining coordinates. The number of sites is here $N=L^d$ on a hypercubic lattice of length $L$. Performing an adiabatic evolution from this Hamiltonian to $\sum_k \varepsilon(k) c^\dagger(k)c(k)$ with some dispersion relation $\varepsilon(k)=\varepsilon(k_1,...,k_d)$, we have again a factorization into one-qubit Hamiltonians corresponding to momenta $k=(k_1,...,k_d)$ and $-k=(-k_1,...,-k_d)$. They read
\begin{equation}
H_k(t)=\varepsilon(k)\frac{I-Z}{2} +[2(1-t/T)\sin k_1] Y\,.
\end{equation}
Similar reasonings as before apply, and we get $T_k=\frac{8\sin k_1}{\pi\varepsilon(k)^2}$, and an energy density in the intermediate regime decaying as
\begin{equation}
    \delta e\sim \iint |\varepsilon(k)|e^{-T/T_k}\D{k_1}...\D{k_d}\,.
\end{equation}
Let us for example assume that $\varepsilon(k)$ vanishes at $k^*$ as $\varepsilon(k)\sim |k-k^*|^z$, and that $k_1^*\neq 0,\pi$. Then we find
\begin{equation}
    \delta e\sim T^{-\frac{z+d}{2z}}\,.
\end{equation}
As for the effect of noise, it is more complicated to describe in higher dimension because it will much depend on which fermionic encoding is used, and is beyond the scope of this work.

\subsubsection{Effect of finite adiabatic time}
Let us finally comment on the effect of finite adiabatic time $T$ on the distribution of momenta in the state prepared, beyond just the value of the energy in \eqref{deltae}. We saw that the probability that modes $k$ and $-k$ are simultaneously wrongly occupied or wrongly empty is $e^{-T/T_k}$ with $T_k$ given in \eqref{tk}. So the state that is prepared for finite $T$ is approximately
\begin{equation}
    \begin{aligned}
|\psi\rangle\propto &\prod_{\substack{k>0\\ \cos k>0}}\left(\sqrt{1-p(k)}+\sqrt{p(k)}c^\dagger(-k)c^\dagger(k)\right)\\
\times &\prod_{\substack{k>0\\ \cos k<0}}\left(\sqrt{p(k)}+\sqrt{1-p(k)}c^\dagger(-k)c^\dagger(k)\right)|0\rangle\,,
    \end{aligned}
\end{equation}
with
\begin{equation}
    p(k)=\exp\left(- \frac{2\pi \cos^2 k}{\sin k}T\right)\,.
\end{equation}
For a thermal state with inverse temperature $\beta$, this probability would be
\begin{equation}
  p(k)=  \frac{e^{-4\beta |\cos k|}}{e^{-4\beta |\cos k|}+e^{4\beta |\cos k|}}\approx e^{-8\beta |\cos k|}\,.
\end{equation}
The state prepared for finite adiabatic time $T$ is thus not exactly a thermal state, because the parametric dependence in $k$ differs. The factor $\cos^2 k/\sin k$ compared to the thermal state has the effect of suppressing fluctuations far from the Fermi point $k=\pi/2$, and on the opposite enhancing fluctuations near the Fermi point, compared to a thermal state.

\subsection{Noisy numerical study}
\subsubsection{Generalities}
\begin{figure}
    \centering
    \includegraphics[width=\linewidth]{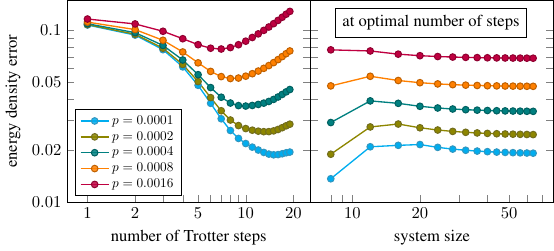}
    \caption{\emph{Left panel:} Numerical simulations for the energy density above ground state as a function of number of Trotter steps in the adiabatic evolution, for different noise rate $p$, in size $N=32$. \emph{Right panel:} Energy density above ground state as a function of number of system size $N$, at the number of Trotter steps that minimizes the energy for a given noise rate $p$. These simulations assume perfect preparation of the initial state \eqref{gs}, which can be in practice always replaced by $|+..+\rangle$ and a post-selection onto even particle number, which does not require any two-qubit gates.}
    \label{fig:noisy_adiabatic}
\end{figure}

We now perform a numerical study of the efficiency of the adiabatic algorithm in the presence of noise. We model the noise by a two-qubit depolarizing channel after every two-qubit gate. We present the results of numerical simulations of the adiabatic evolution in Fig \ref{fig:noisy_adiabatic}, performed with Pauli string simulations. In the left panel we plot the energy density above ground state for a fixed system size $N=32$, as a function of the number of Trotter steps in the noisy adiabatic evolution. We see that the curves generically decrease up to some optimal number of steps $T^*$, and then increase because of hardware noise. As expected, $T^*$ decreases with the noise rate $p$. In the right panel, we then plot the optimal energy density obtained at $T^*(p)$ with the noisy adiabatic evolution, as a function of system size $N$. We see that this energy density is constant with system size at large $N$, and from the logarithmic scale we can see that it scales approximately as $\sqrt{p}$, in agreement with formula \eqref{sqrtp}.

\subsection{Classically-optimized parametrized circuits (COP)}
A popular way to prepare low-energy states on current hardware where resources are limited is to use parametrized circuits that contain a fixed number of gates, where the parameters are classically optimized so as to minimize the energy. This approach has been implemented on hardware in several recent works \cite{herrmann2022realizing,lin2026utility,shen2026realization,granet2025superconducting}. It relies on the classical simulability of the state preparation circuits, and is thus limited for general Hamiltonians to either small number of qubits (for statevector simulation) or shallow circuits (for tensor network \cite{robertson2025approximate} or Pauli string simulations \cite{lin2026backpropagating}). However, for state preparation with FFT, the states obtained are typically going to be Gaussian states (for example, if it is a product state in momentum basis). Such states can thus in principle be obtained with only Gaussian, free-fermion circuits that can be simulated efficiently for any system size. In the case of ground state  preparation for example, a natural adiabatic-inspired ansatz is
\begin{equation}\label{vqe}
\begin{aligned}
&|\psi(x_1,...,x_M,y_1,...,y_M)\rangle=\\
&\prod_{m=1}^{M} e^{i x_m\sum_{j=1}^N X_j X_{j+1}}e^{i y_m \sum_{j=1}^N Y_j Y_{j+1}} \frac{|+...+\rangle+|-...-\rangle}{\sqrt{2}}\,.
\end{aligned}
\end{equation}
It corresponds to taking the same circuit structure as for an adiabatic evolution, but now taking all the gate angles as parameters to vary. In the case of free fermions, these COP states are legitimate because their optimization is scalable (polynomial in system size and circuit depth) and can be done for large system sizes \cite{dreyer2021quantum}.

\begin{figure*}
    \centering
    \includegraphics[width=\linewidth]{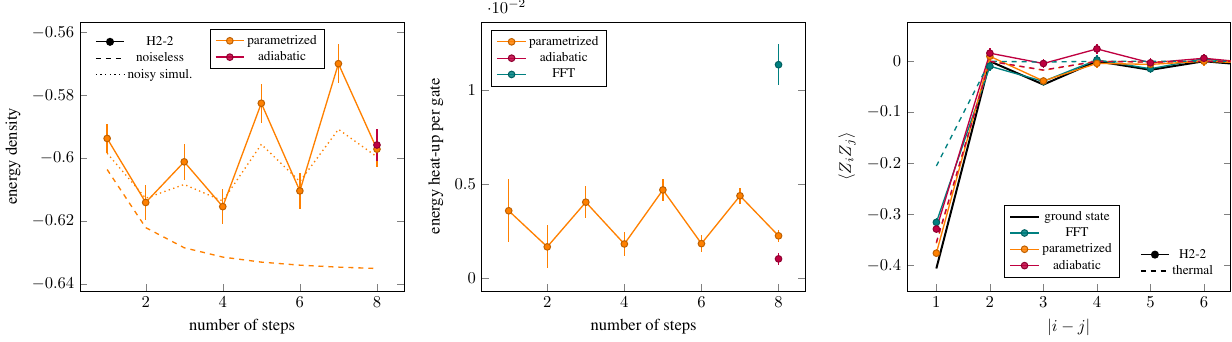}
    \caption{\emph{Left panel:} Energy density as a function of number of steps in the ansatz of the parametrized circuit, in size $N=48$, measured on Quantinuum H2-2 quantum computer (solid lines), noiseless values (dashed lines) and noisy simulations with depolarizing noise $0.001$ after every two-qubit gate (dotted lines). We indicate the value obtained with an adiabatic evolution for $8$ steps as a comparison. \emph{Middle panel:} Energy increase per two-qubit gate compared to noiseless due to noise, as a function of number of steps, measured on Quantinuum H2-2 quantum computer, from the same data as left panel. We also plot as a comparison the FFT value from Fig \ref{fig:energy_comparison}, with a number of two-qubit gates roughly corresponding to $8$ steps. \emph{Right panel:} Correlation function $\langle Z_i Z_j\rangle$ as a function of distance $i-j$, comparing the different methods on hardware (bullets), as well as the value obtained for a thermal state with same energy (dashed), and exact correlations in the ground state (solid). The thermal curves for parametrized and adiabatic are almost superimposed.}
    \label{fig:vqe_energy}
\end{figure*}

In practice, we observe significant improvement of the energies that can be obtained with COP compared to adiabatic, with the same number of gates. Efficiency has similar features as the free-fermion model previously studied in Ref \cite{dreyer2021quantum}. In particular, the ground state can be prepared exactly with $M=N/2$ steps.  In Fig \ref{fig:vqe_energy}, we plot the energies obtained with COP circuits in size $N=48$, comparing different number of steps in the ansatz, when run on Quantinuum H2-2 hardware. We observe that the lowest energy is obtained for $4$ steps, with an energy density $-0.615 \pm 0.005$, exact being $-0.637$, which is the lowest energy obtained on this system size by any method in this paper. For a thermal state, this energy density corresponds to an inverse temperature $\beta=3.65$. The adiabatic evolution with $8$ steps turns out to give very similar energies on hardware as the parametrized state, with the same number of gates. 

In the middle panel of Fig \ref{fig:vqe_energy}, we observe a striking dependence of sensitivity to noise on the number of Trotter steps, with a heat-up per gate that is almost twice larger for odd number of steps. This effect is also observed in noisy numerical simulations where a depolarizing noise channel is added after every two-qubit gate. It is not observed for noisy simulations of the adiabatic evolution, which implies that it must come from the particular pattern of the gate angles, as the structure of the circuit is otherwise identical. It turns out that the optimal gate angles for even number of Trotter steps are more regular in time than for odd number of Trotter steps, see Appendix \ref{cop}. We also observe that although not specific to the hardware, the amplitude of the effect is significantly larger on hardware compared to noisy simulations. We have checked that including a gate-angle dependence in the amplitude of the noise (which better describes noise on Quantinuum hardware, see Ref \cite{moses2023race}) does not significantly modify the noisy simulation curve and cannot explain alone the more pronounced effect on hardware. The precise origin of this effect and its  enhancement on hardware is thus unclear. It is in any case further demonstration that the structure of the circuit can significantly influence sensitivity to noise, beyond the number of gates and the depth.

Apart from the even/odd variations, the heating per gate for the parametrized state is roughly constant with the number of steps, at fixed system size. This is in agreement with dilution of error \cite{granet2025dilution}. Indeed, in the absence of dilution of error, the number of errors that are within the lightcone in one dimension grows quadratically with the number of steps, whereas the number of gates grows linearly, which should result in a heating per gate that grows linearly with the number of steps. In terms of amplitude, the heating per gate for the parametrized state is around twice larger than for the adiabatic evolution, and around three times smaller than for a FFT.

Finally, in the right panel of Fig \ref{fig:vqe_energy}, we plot the correlation $\langle Z_i Z_j\rangle$ in the state prepared in size $N=48$ measured on hardware, comparing the different approaches. We also compare them to the ground state correlations, and to the correlations within the thermal state with same energy. We observe that the parametrized state reproduces the best the ground state correlations, although differing from the thermal value. We also note that the correlations within the FFT state beyond distance $2$ are close to the ground state values, although the energy is significantly higher.

\subsection{Relation between noise sensitivity and precision in momentum space}
We now argue for a relation between sensitivity to noise and precision in momentum space. As said above, the FFT is a fermionic operator that is defined by its action of the fermionic modes
\begin{equation}
    {\rm FFT}^\dagger c_j {\rm FFT} = \sum_{\ell=0}^{N-1}c_\ell F_{\ell j}\,,\qquad F_{\ell j} =\frac{e^{\frac{2i\pi}{N}j\ell}}{\sqrt{N}}\,.
\end{equation}
Contrary to elsewhere in the manuscript, for convenience and simplicity of notations we have taken periodic boundary conditions and labeled the modes from $0$ to $N-1$. It is a Gaussian fermionic operator, and so can be written as an exponential of a two-body Hamiltonian $H$, i.e.
\begin{equation}
    {\rm FFT}=\exp \left(i\sum_{j,\ell=0}^{N-1} h_{j\ell}c^\dagger _j c_\ell \right)\,,\qquad ih=\log F\,.
\end{equation}
Since $(F^2)_{j\ell}=\delta_{j,N-\ell}$ where the indices are understood modulo $N$, we have $F^4=I$, and so $h$ can be expressed as a linear combination of $I,F,F^2$ and $F^3$. Replacing $F$ by $1,i,-1$ and $-i$, which satisfy the same algebra, and defining the logarithm branch as $\log e^{i\theta}=i\theta$ for $-\pi < \theta \leq \pi$, we can identify the coefficients. It yields
\begin{equation}
   h=\frac{\pi}{4}(I+F^2-(1+i)F-(1-i)F^3)\,,
\end{equation}
namely, we obtain ${\rm FFT}=e^{iH_{\rm FFT}}$ with the Hamiltonian
\begin{equation}
\begin{aligned}
    H_{\rm FFT}=&\frac{\pi}{\sqrt{2N}}\sum_{j,\ell=0}^{N-1} c_j^\dagger c_\ell \sin \left(\frac{2\pi}{N}j\ell-\frac{\pi}{4}\right)\\
    &+\frac{\pi}{4}\sum_{j=0}^{N-1}[c_j^\dagger c_{N-j}+c_j^\dagger c_j]\,.
\end{aligned}
\end{equation}
We note that because of an arbitrary choice of logarithm branch, there is no unique Hamiltonian that exponentiates to the FFT. We see that this Hamiltonian whose time evolution is the FFT is completely all-to-all coupled. The coefficient of the two-fermion term does not decay with distance and is oscillatory.

\begin{figure}
    \centering
    \includegraphics[width=0.6\linewidth]{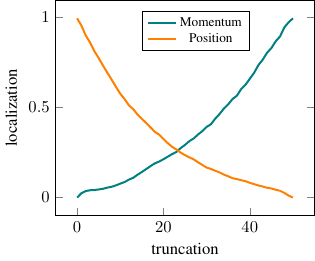}
    \caption{Standard deviation of $|\alpha^{(n)}_{j,\ell}|^2$ (teal) and $|\beta^{(n)}_{j,\ell}|^2$ (orange) corresponding to $\ell$, multiplied by $\sqrt{N}$ for normalization, and averaged over $j$, for a system size $N=100$, as a function of truncation parameter $n$.}
    \label{fig:locfft}
\end{figure}

Let us instead consider the following truncation of the FFT Hamiltonian for an integer $n\geq 1$
\begin{equation}
\begin{aligned}
    H_{\rm FFT}^{(n)}=&\frac{\pi}{\sqrt{2N}}\sum_{\substack{j,\ell=0\\ d(j,\ell)\leq n}}^{N-1} c_j^\dagger c_\ell \sin \left(\frac{2\pi}{N}j\ell-\frac{\pi}{4}\right)\\
    &+\frac{\pi}{4}\sum_{\substack{j=0\\ d(j,N-j)\leq n}}^{N-1}c_j^\dagger c_{N-j}+\frac{\pi}{4}\sum_{j=0}^{N-1}c_j^\dagger c_j\,,
\end{aligned}
\end{equation}
with the distance $d(j,\ell)=\min(|j-\ell|,|j+N-\ell|,|j-N-\ell|)$ that takes the periodic boundaries into account. This corresponds to $H_{\rm FFT}$ with all terms separated by a distance $>n$ truncated. For $n=N/2$, we have $H^{(N/2)}_{\rm FFT}=H_{\rm FFT}$, while for lower values of $n$, $H_{\rm FFT}^{(n)}$ is increasingly more local. We define $\alpha_{j\ell}^{(n)},\beta_{j\ell}^{(n)}$ by
\begin{equation}
\begin{aligned}
    e^{-i H^{(n)}_{\rm FFT}}c_j e^{i H^{(n)}_{\rm FFT}}&=\sum_{\ell=0}^{N-1}\alpha^{(n)}_{j\ell} c(2\pi \ell/N)\\
    &=\sum_{\ell=0}^{N-1}\beta^{(n)}_{j\ell} c_\ell\,,
\end{aligned}
\end{equation}
where we recall that $c(k)$ is the Fourier transform of $c_j$, namely $c(k)=\frac{1}{\sqrt{N}}\sum_{j=0}^{N-1}e^{2i\pi jk/N}$. The quantity $\alpha$ (resp. $\beta$) gives information about how much the application of $e^{i H^{(n)}_{\rm FFT}}$ on $c_j$ gives an operator that is localized in momentum space (resp. real space). For example for the FFT, we have $\alpha_{j\ell}^{(N/2)}=\delta_{j,\ell}$, namely, the FFT of $c_j$ is perfectly localized in momentum space. For $n=0$, we have $e^{-i H^{(n)}_{\rm FFT}}c_j e^{i H^{(n)}_{\rm FFT}}=c_j$ and so $\alpha_{j\ell}^{(N/2)}=\frac{e^{-2i\pi j\ell/N}}{\sqrt{N}}$ which is completely delocalized in momentum space. For the quantity $\beta$, these two limit cases are reversed. In order to quantify the spread in momentum and real space, we plot in Fig \ref{fig:locfft} the standard deviation with $\ell$ of $|\alpha^{(n)}_{j,\ell}|^2$ and $|\beta^{(n)}_{j,\ell}|^2$, at a fixed $j$, and then average this variance over $j$. This measures how much the $\alpha$'s and $\beta$'s have a uniform amplitude, namely how much the operators are localized in real space or momentum space. We observe that as we take $n$ smaller, namely as the Hamiltonian $H^{(n)}_{\rm FFT}$ is more local in space, the resulting operators are less precise in momentum space and more precise in real space.

Now, long-range hoppings $c_j^\dagger c_{j+\delta}+c_{j+\delta}^\dagger c_{j}$ are more costly to implement and more sensitive to noise as $\delta$ grows, because of the fermionic anticommutation relations. Let us explain this for a Jordan-Wigner transformation, where this operator is expressed as $X_jZ_{j+1}...Z_{j+\delta -1}X_{j+\delta}$ plus the same operator with $Y$ instead of $X$. These long $Z$-strings require some form of gadgets that will propagate every error on multiple qubits. Let us for example consider the case of a single-site depolarizing channel applied on the two qubits of a two-qubit gate. After a $X_1X_2$ rotation, there are $6$ equally likely errors, $X_1$, $Y_1$, $Z_1$, $X_2$, $Y_2$, $Z_2$. Now, when implementing a hopping for $\delta=2$, one must implement a rotation of $X_1Z_2X_3$. To express it in terms of two-qubit gates, one must use a gadget such as ${\rm CZ}_{12} e^{i\theta X_{1}X_3} {\rm CZ}_{12}$. There are now $3$ two-qubit gates, so $3$ times more errors. But also, many of these errors will  spread onto multiple qubits. The effective errors of the first ${\rm CZ}_{12}$ become $X_1Z_2$, $e^{-2i\theta X_1X_3}Y_1 Z_2$, $e^{-2i\theta X_1X_3}Z_1$, $Z_1X_2$, $Z_1Y_2$, $Z_2$, those of the $XX$ rotation become $X_1Z_2$, $Y_1Z_2$, $Z_1$, $X_3$, $Y_3$, $Z_3$, and those of the second ${\rm CZ}_{12}$ are unchanged and remain $X_1$, $Y_1$, $Z_1$, $X_2$, $Y_2$, $Z_2$. Even without counting the $X_1X_3$ rotations appearing in the errors (that will propagate errors more slowly than Pauli matrices for small $\theta)$, each error spreads on average over $1.28$ qubits. At the end of the circuits, these errors will thus be more likely to impact more qubits, and have a larger effect on expectation values of local observables. For larger $\delta$, one needs larger gadgets, which leads to even more spreading. Not only the number of two-qubit gates to implement a long-range hopping is larger (which is by itself not related to error sensitivity), but also every two-qubit gate will in average create errors on more qubits, which generally implies that the heat-up per gate will be higher. These arguments show thus a connection between the precision of the information we can have in momentum space, and the sensitivity to noise of the circuits that implement the transformation.

\section{Measuring momentum with local circuits}
\subsection{Setup and objective}
We now present another application where the FFT is replaced by a less precise but also less costly and less noisy alternative. 

Beyond state preparation, the FFT can also be used in the reverse way to measure momentum modes in the state prepared. In that case, by measuring in the $Z$ basis every qubit after doing a FFT, one obtains snapshots of mode occupation in momentum space. This can be used for example to measure the expectation value of $n(k)\equiv c^\dagger(k)c(k)$ the average density of momentum $k$. However, for physical systems, this is often too precise information, as two neighbouring modes separated by $2\pi/N$ will contain equivalent physical information for large systems. For many practical purposes, it would suffice to be able to detect momenta within a certain window $[k,k+\Delta k]$ that is independent of $N$. 

\subsection{Momentum density profile reconstruction from local measurements}

As an alternative to the FFT to measure the momentum distribution of a state, we propose the following method that we will refer to as MDLM (Momentum Distribution from Local Measurements). Let us consider the operators
\begin{equation}
\begin{aligned}
    O_n&=\frac{1}{2}\sum_{j=1}^N c^\dagger_j c_{j+n}+c^\dagger_{j+n} c_{j}\\
    O_n'&=\frac{i}{2}\sum_{j=1}^Nc^\dagger_j c_{j+n}-c^\dagger_{j+n} c_{j}\\
\end{aligned}
\end{equation}
for $n\geq 0$. Before explaining how these operators can help us measuring momentum modes, let us first explain how they can be measured efficiently on hardware. We spell out a gadget $F_n$ for measuring $O_n$ described in Algo 1. It is written in the case of $N$ even, and assuming for simplicity that $N$ is a multiple of $n$ when $n$ is even.  The operator $F_{n}$ has the property that
\begin{equation}
    F_n c_{2j}^\dagger c_{2j+1}F_n^\dagger= c^\dagger_{x_j}c_{x_j+n}
\end{equation}
for a certain injective function $j\mapsto x_j$. After applying $F_n$, fermions separated by one site, sitting on sites $2j$ and $2j+1$, are now separated by $n$ sites. The formula however only applies to $c^\dagger_{2j}c_{2j+1}$, and not $c^\dagger_{2j+1}c_{2j+2}$, which yields only half of the pairs of fermions separated by $n$ sites. The remaining half can be recovered by shifting all sites by $1$ site when $n$ is odd, and by shifting all sites by $2q+2$ sites when $n$ is even.

\begin{figure}
    \centering
    \includegraphics[width=\linewidth]{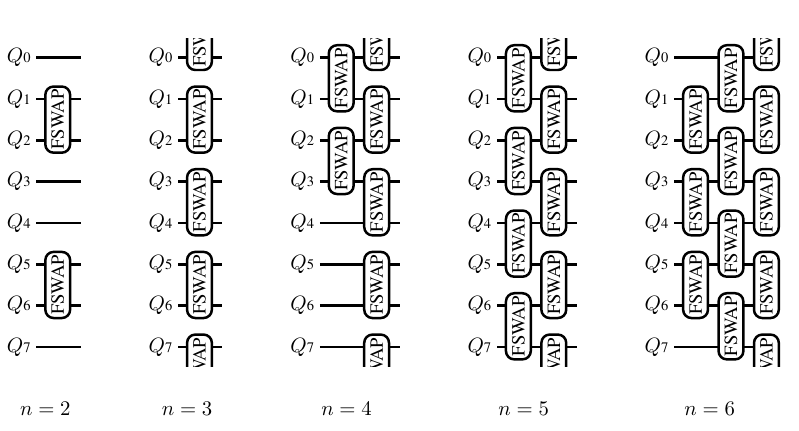}
    \caption{Illustration of the FSWAP circuit applied on the qubits on the left, generated by the measurement gadget given in Algo 1 for different values of $n$.}
    \label{fig:brickwall}
\end{figure}

This operator $F_n$ is defined in terms of fermionic swaps (FSWAP) that are $e^{i\pi/2 (c_i^\dagger c_j+c_j^\dagger c_i)}$ between fermionic modes $i$ and $j$. In the Jordan-Wigner encoding, when $i$ and $j$ are neighbours, the FSWAP is a swap followed by a CZ gate. The operator $F_n$ is decomposed as a simple brickwall of FSWAPs when $n$ is odd, as illustrated in Fig \ref{fig:brickwall}. When $n$ is even, this brickwall of FSWAPs has to be supplemented with a row of FSWAPs alternating $n/2$ adjacent FSWAPs and $n$ qubits untouched. By applying $F_n$ on the qubit register, measuring $O_n$ is reduced to measuring $O_1$. 

The cost of implementing this measurement gadget is a depth $\lfloor n/2\rfloor$ and $(n-1)N/4$ FSWAPs for $n\geq 1$. It is thus relatively cheap to implement, both in number of gates and in circuit depth. We note that there exist other even cheaper ways of measuring these operators, for example by measuring in the $X$ and $Z$ basis appropriate sets of qubits. However, the advantage of the approach using the  gadget $F_n$ is that one can measure the number of particles simultaneously with $O_n$.

\begin{figure}
    \centering
    \includegraphics[width=\linewidth]{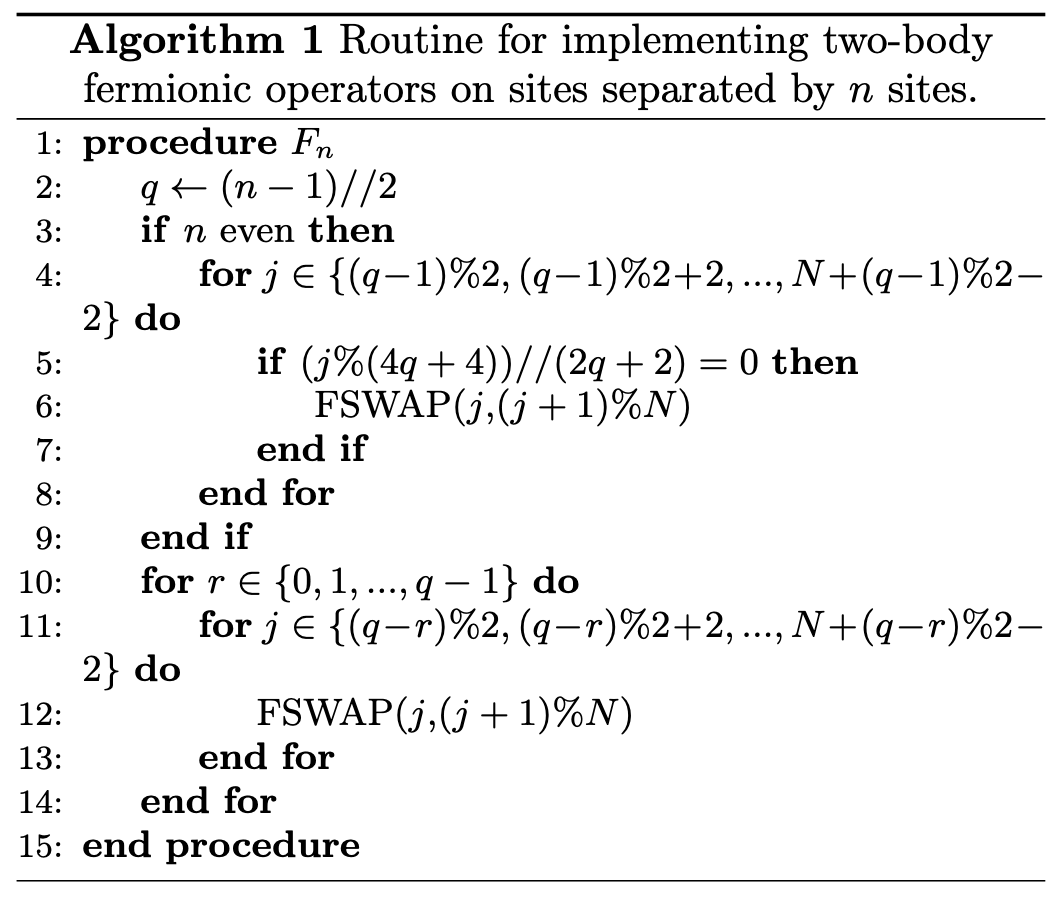}
    \label{algo_1}
\end{figure}


Let us now explain how the operators $O_n$ can be used to deduce momentum mode occupation. The expectation value of $O_n$ can be written in terms of the momentum mode occupation numbers $n(k)$ as
\begin{equation}
\begin{aligned}
    \langle O_n\rangle&=\sum_{k}\cos(kn) \langle n(k)\rangle\\
    \langle O'_n\rangle&=\sum_{k}\sin(kn) \langle n(k)\rangle\,.
\end{aligned}
\end{equation}
The measurement of these observables gives thus information on $\langle n(k)\rangle$, as it puts constraints on the values that the set $\{\langle n(k)\rangle\}$ can take. Schematically, when measuring $M$ of these observables, there remain $N-M$ degrees of freedom in the momentum densities $\langle n(k)\rangle$. In order to fix these degrees of freedom while inputting as little extra information as possible, one maximizes the Shannon entropy
\begin{equation}
    S(\{p\})=-\sum_{k} [p_k \log p_k+(1-p_k)\log (1-p_k)]\,,
\end{equation}
with $p_k\equiv \langle n(k)\rangle$, at the fixed constraints imposed by the measured value of $\langle O_n\rangle$ and $\langle O'_n\rangle$. This constrained optimization problem can be expressed with Lagrange multipliers $\beta_n,\beta'_n$, and looking for stationary points of
\begin{equation}\label{lagrange}
\begin{aligned}
    \mathcal{L}(\{\beta,\beta',p\})=S(\{p\})&-\sum_{n=0}^M \beta_n \left(\langle O_n\rangle-\sum_k \varepsilon_{n,k}p_k\right)\\
    &-\sum_{n=1}^M \beta_n' \left(\langle O_n'\rangle-\sum_k \varepsilon_{n,k}'p_k\right)\,,
\end{aligned}
\end{equation}
where we have set $\varepsilon_{n,k}=\cos(kn)$ and $\varepsilon_{n,k}'=\sin(kn)$. When writing $\partial_{p_k}\mathcal{L}=0$, we obtain
\begin{equation}\label{p_k}
    p_k=\frac{1}{1+e^{-\sum_{n=0}^M \beta_n \varepsilon_{n,k}-\sum_{n=1}^M\beta'_n \varepsilon_{n,k}'}}\,.
\end{equation}
Since $\partial_{p_i}^2\mathcal{L}(\{\beta,\beta',p\})=-\frac{1}{p_i(1-p_i)}<0$ and all cross-derivatives vanish, \eqref{p_k} is a local maximum of $\mathcal{L}$ at fixed $\beta_n$. Then the cost function becomes
\begin{equation}
\begin{aligned}
    \mathcal{L}(\{\beta,\beta'\})=&\sum_{k}\log\left(1+e^{\sum_{n=0}^M \beta_n \varepsilon_{n,k}+\sum_{n=1}^M\beta_n' \varepsilon_{n,k}'}\right)\\
    &-\sum_{n=1}^M \beta_n \langle O_n\rangle+\beta'_n \langle O'_n\rangle\,.
\end{aligned}
\end{equation}
One must now minimize this value $\mathcal{L}(\{\beta,\beta'\})$ so as to maximize the entropy at fixed values of constraints.

\begin{figure*}
    \centering
    \includegraphics[width=\linewidth]{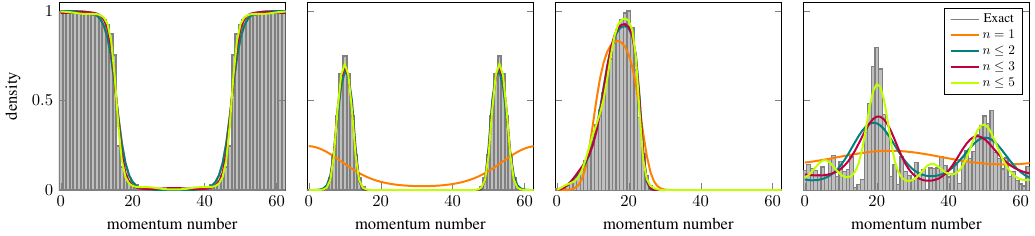}
    \caption{\emph{Numerical results.} Momentum density in four different numerical settings in size $N=64$, comparing exact (gray bars) with reconstructed profiles (lines) with the MDLM protocol, including observables $O_n$ and $O'_n$ with $n=1$, $n=1,2$, $n=1,2,3$ and all $n\leq 5$.}
    \label{fig:mdlm}
\end{figure*}

In practice, the measured values of $\langle O_n\rangle$ are going to come with some error bars $\sigma_n$ when measured on a quantum computer, that should translate into some error bars on the estimated $p_k$. We will estimate these error bars the following way. We generate random values $\tilde{\langle O_n\rangle}$ by adding to $\langle O_n\rangle$ a random Gaussian variable with mean $0$ and standard deviation $\sigma_n$. Then we compute the maximal entropy value of $p_k$ satisfying the constraints given by $\tilde{\langle O_n\rangle}$. By repeating this process several times we can estimate a variance on the resulting $p_k$. We note that adding random values to the measured $\langle O_n\rangle$ can in principle gives a set of $\langle O_n\rangle$ that cannot be obtained with a set of $p_k$ satisfying $0\leq p_k\leq 1$ for all $k$. However, this fact is also true for shot noise coming from running circuits on hardware: it is already possible that a set of expectation values $\langle O_n\rangle$ computed with a finite number of shots cannot be obtained with all $0\leq p_k\leq 1$.

When restricting to only $O_n$ with $n=1$, the output distribution for $\langle n(k)\rangle$ is that of a thermal state of a free-fermion hopping chain with energy density matching the energy density measured with $O_n$. When taking into account $n>1$ and $O'_n$, more general distributions can be recovered. These distributions will allow for faster variations in momentum space and for a larger number of minima/maxima. There is again a correspondence between the locality of the observable measured in real space and the amount of variations allowed in momentum space. Recovering sharp or multiple variations in momentum space requires to measure circuits that are less local, which comes with more gates and propagation of errors, and so is more sensitive to hardware noise.

\subsection{Numerical tests}
We present in Fig \ref{fig:mdlm} some numerical tests of our protocol. We consider four different states in size $N=64$, from left to right, that are taken to be product states in momentum space with different densities. The first state is similar to an approximate ground state preparation of a free-fermion hopping chain, the second state is similar to typical outcomes of spectral function measurement, see Ref \cite{granet2026spectral} and below, the third state is an arbitrary non-symmetric wave packet in momentum space, and the fourth state some noisy random data with two peaks. We compare the exact momentum densities with those obtained with MDLM, taking into account $\langle O_n \rangle$ and $\langle O'_n \rangle$ for different values of $n$, and neglecting shot noise. We observe that in all settings, including higher values of $n$ up to $n=5$ always improves the agreement with the exact value. When including only $n=1,2$, we already observe very good agreement between the reconstructed momentum profile and the exact profile for the first three cases where the exact profile is smooth. For the fourth noisy case, the MDLM is not precise enough to recover each variation from one momentum to the next, but is able to recover the overall shape of the momentum density profile.

\begin{figure*}
    \centering
    \includegraphics[width=\linewidth]{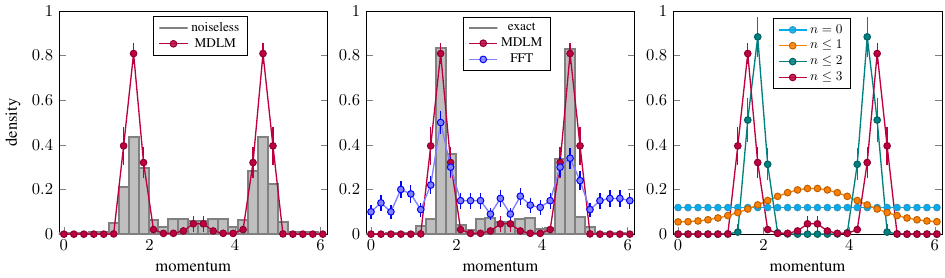}
    \caption{\emph{Hardware results.} Momentum density obtained with adiabatic evolution and MDLM, using $n=0,1,2,3$ measured on Quantinuum H2-2 quantum computer (first two panels), comparing with noiseless and with exact. The FFT curve is a Quantinuum H2-2 quantum computer result taken from Ref \cite{granet2026spectral} where FFT is used both for state preparation and measurement. The third panel shows the convergence of the reconstructed momentum density as more $\langle O_n\rangle$ are taken into account.}
    \label{fig:hardware}
\end{figure*}
\subsection{Hardware experiment}
We now present a hardware experiment that incorporates both the adiabatic state preparation of Section \ref{sec:noisesensitive} and our MDLM protocol for measuring momentum density. The setup that we consider is that of Ref \cite{granet2026spectral}. There, a quantum circuit was proposed to measure the so-called spectral function $A(k,\omega)$ of a condensed-matter system, for arbitrary momenta $k$ and frequency $\omega$. The quantum circuit involves the simulation of the coupling between a system and an environment, defined on the same number of qubits as the system. While the frequency $\omega$ is hard-coded into the total Hamiltonian, the different momenta $k$ are obtained from measuring the momentum density $n(k)$ in the environment. In Ref \cite{granet2026spectral}, this was done with a FFT. Here, we replace this FFT operation by our MDLM protocol, measuring only even observables $O_n$ for $n=0,1,2,3$. Moreover, in Ref \cite{granet2026spectral}, the initial state preparation was done with a FFT. We replace it with an adiabatic evolution, doing $7$ steps. Apart from state preparation and measurement, the remainder of the circuits is identical. We performed $300$ shots for $O_0,O_2,O_3$ and $375$ shots for $O_1$.

We present the hardware results in Table \ref{fig:os} for the values of $O_n$ and in Fig \ref{fig:hardware} for the reconstructed densities. We compare the hardware results to two momentum density profiles, "noiseless" and "exact". "Noiseless" denotes the density profile that we would obtain from the circuits if they were noiseless and if all observables $O_n$ for $n=0,1,...,N-1$ were measured. Equivalently, this is what we would obtain if we did a FFT instead of the MDLM to measure the density profile, and assumed noiseless hardware. "Exact" denotes the density profile obtained if the state preparation and measurement are both perfect and noiseless, namely if both are done with a FFT on perfect hardware. It is identical data to what was computed in Ref \cite{granet2026spectral}. We also compare our results to what was measured in Ref \cite{granet2026spectral} with the FFT for both state preparation and momentum density measurements.

We observe very good agreement between the exact momentum profile and what is reconstructed with MDLM, taking into account just $O_n$ up to $n=3$. In particular, we see that MDLM correctly finds zero momentum density for the values near $0$ and $2\pi$, and has no noise floor compared to the noisy FFT. The peaks are perfectly located and the amplitude relatively well reproduced. Compared to noiseless, the peaks are again perfectly located, but the reconstructed amplitude of the peaks is larger.

\begin{table}
\centering
\begin{tabular}{|c|c|}
\hline
    $O_0$ & $0.11914 \pm 0.00352$ \\
\hline
    $O_1$ & $0.03713  \pm 0.00623$ \\
\hline
    $O_2$ & $0.10417   \pm  0.00742 $ \\
\hline
    $O_3$ & $-0.01194 \pm 0.00698  $ \\
\hline
\end{tabular}
    \caption{Expectation values of $O_n$'s in the environment in the setup described in the text, measured on Quantinuum H2-2 quantum computer.}
    \label{fig:os}
\end{table}

\section{Discussion and outlook}
This work originated from the observation that the FFT on actual quantum computer is more noisy than typical circuits used for condensed-matter physics or physical simulations in general, as implemented in Ref \cite{granet2026spectral}. We demonstrate on Quantinuum H2-2 quantum computer that when preparing the ground state of a free-fermion model, single errors in FFT impact the ground state energy density by $\mathcal{O}(1)$, whereas single errors in adiabatic evolution impact the ground state energy density by only $\mathcal{O}(1/N)$. In particular, there is a critical system size $N\approx 20$ beyond which an adiabatic evolution reaches lower energies than the FFT, \emph{with the same number of gates}. This emphasizes the importance of decreasing the \emph{noise sensitivity} of quantum algorithms, beyond the number of gates or circuit depth.

This noise sensitivity is related to how errors propagate in the circuit. The difference between FFT and standard routines such as Hamiltonian simulation or adiabatic evolution is that the FFT is a "global" circuit that connects qubits far apart with a lot of maximally-entangling Clifford gates, whereas Hamiltonian simulation involves only neighbouring qubits on a given geometry with typically small-angle rotation gates. These long-range gates (in real space) in the FFT reflect the high precision that FFT gives in momentum space, as it can distinguish momenta separated by $2\pi/N$. The price to pay for this high precision is high sensitivity to errors. For physical applications however, this high precision is not required, as only a precision $\mathcal{O}(1)$ on momentum is typically required. 

We have then presented alternatives to FFT that are \emph{less precise} but also \emph{less noisy}. For physical applications of FFT such as state preparation or momentum measurement, this loss of precision has low impact on physically relevant quantities, while the reduction of noise sensitivity drastically improves the results on actual hardware. We demonstrated on Quantinuum H2-2 quantum computer the measurement of spectral functions with a technique utilizing only local circuits to measure momentum (at the price of finite precision) and obtained significantly better results than with the FFT, recovering the position of the peaks of the spectral function correctly. Using classically-optimized parametrized circuits, we obtained in the best case the preparation of a low-energy state of a tight-binding chain on $48$ qubits, with an energy corresponding to that of a thermal state with an inverse temperature $\beta=3.65$.

This work leaves open a number of questions. For example, in this work we only discussed the use of the adiabatic evolution for preparing ground states. However, the FFT can also prepare arbitrary excited states, and it would be relevant to ask whether an adiabatic protocol with similar low noise sensitivity can be devised for these excited states. Another open direction is whether the FFT could be modified in a smooth way (instead of completely changing the circuit into an adiabatic evolution circuit) so as to smoothly decrease noise sensitivity, at the cost of smoothly decreasing momentum precision. A "quantum wavelet transform", where the target states have a finite width in momentum space and are partly localized in space could potentially fulfill this objective.

\section*{Acknowledgements}
We thank Andrew Potter and Sheng-Hsuan Lin for helpful comments on the manuscript. The experimental data reported in this work were produced by the Quantinuum H2-2 quantum computer, Powered by Honeywell, in June-July 2026. 


%

\onecolumngrid
\appendix
\section{Hardware gate counts \label{appgatecount}}
We give in the following table the two-qubit gate counts and number of adiabatic steps for the hardware experiments run for Fig \ref{fig:energy_comparison}. These include as well the $N/2$ two-qubit gates used for the energy measurement at the end of the circuit, which are identical for both FFT and adiabatic.

\begin{center}
    
   \begin{tabular}{|c|c|c|c|c|c|c|}
\hline
   System size  &  $8$  & $12$ & $16$ & $24$ & $32$ & $48$  \\
\hline
\hline
  FFT two-qubit gate count  &  $49$  & $98$  & $157$ & $296$ & $455$ &  $820$ \\
\hline
  Adiabatic two-qubit gate count  &  $56$  & $108$ & $176$ & $312$ & $480$ &  $912$ \\
\hline
  Number of adiabatic steps  &  $3$  & $4$ & $5$ & $6$ & $7$ &  $9$ \\
\hline
\end{tabular}
\end{center}

\section{Optimal angles of COP circuits \label{cop}}
\begin{figure}
    \centering
    \includegraphics[width=0.7\linewidth]{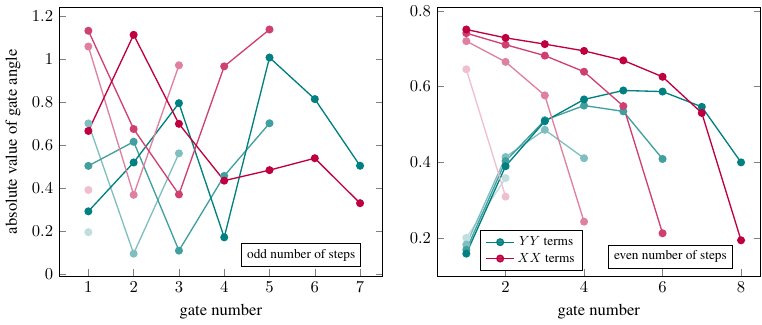}
    \caption{Absolute value of gate angles entering the optimal COP circuits in size $N=48$, for different number of steps from $1$ to $8$, with odd number on left panel and even numbers on right panel, with lighter representing the smallest numbers and darker the highest numbers.}
    \label{fig:gate_angles}
\end{figure}

We display here the optimal angles obtained for the COP circuits in size $N=48$, as a function of the number of steps. The parameters in \eqref{vqe} are $x_t=d_t/2$ and $y_t=\alpha_t d_t/2$. The absolute value of the gate angles are displayed in Fig \ref{fig:gate_angles}, separating even and odd number of steps, to emphasize the smoothness for even number of steps and apparent randomness for odd number of steps.

\begin{center}
\begin{tabular}{|c|c|c|c|c|c|c|c|c|}
\hline
 & $N_{\rm steps}=1$ &$N_{\rm steps}=2$ &$N_{\rm steps}=3$ &$N_{\rm steps}=4$ &$N_{\rm steps}=5$ &$N_{\rm steps}=6$ &$N_{\rm steps}=7$ &$N_{\rm steps}=8$\\
\hline
  $\alpha_1$   & 0.49999976& 0.31003557& 0.66259056& 0.25329997& 0.44553430& 0.22740123& 0.43987742& 0.21104073 \\
\hline
  $d_1$   & 0.78539801& 1.29125894& 2.12031284& 1.44023090& 2.26677215& 1.48207765& 1.33475615& 1.50162366 \\
\hline
  $\alpha_2$   & 0& 1.15886068& 0.26100721& 0.62206304& 0.91246201& 0.56714975& 0.46725092& 0.53457958 \\
\hline
  $d_2$   & 0& 0.61827772& 0.74080772& 1.33040437& 1.35267579& 1.42123676 & 2.22914488& 1.45747341 \\
\hline
  $\alpha_3$   & 0& 0& 0.57900506& 0.84239041& 0.29773095& 0.74976734 & 1.13714735& 0.71381754 \\
\hline
  $d_3$   & 0& 0& -1.94561310& 1.15333042& -0.74407743& 1.36354021 & 1.40095014& 1.42431565 \\
\hline
  $\alpha_4$   & 0& 0& 0& 1.69053842 & 0.47357779& 0.86007105 & 0.39574342& 0.81514539 \\
\hline
  $d_4$   & 0& 0& 0& 0.48578925 & 1.93562006& 1.27843672 & -0.87231655 & 1.38861038 \\
\hline
  $\alpha_5$   & 0& 0& 0& 0 & 0.61712696& 0.97447985 & 2.07976530 & 0.88121147 \\
\hline
  $d_5$   & 0& 0& 0& 0 & -2.27862267& 1.09685692 & -0.97015689 & 1.33830859  \\
\hline
  $\alpha_6$   & 0& 0& 0& 0 & 0& 1.92528863 & 1.50919897 & 0.93757259 \\
\hline
  $d_6$   & 0& 0& 0& 0 & 0& 0.42452896 & 1.08092885 & 1.25166327 \\
\hline
  $\alpha_7$   & 0& 0& 0& 0 & 0& 0 & 1.52563605 & 1.03059823 \\
\hline
  $d_7$   & 0& 0& 0& 0 & 0& 0 & 0.66272181  & 1.06035409 \\
\hline
  $\alpha_8$   & 0& 0& 0& 0 & 0& 0 & 0  & 2.06295578 \\
\hline
  $d_8$   & 0& 0& 0& 0 & 0& 0 & 0  & 0.38749358 \\
\hline
\end{tabular}
\end{center}

\end{document}